\documentclass[%
 reprint,
 amsmath,amssymb,
 aps,
]{revtex4-2}

\usepackage{graphicx}
\usepackage{dcolumn}
\usepackage{bm}
\usepackage{orcidlink} 

\begin{document}
\preprint{TDV}

\title{A feasible dose–volume estimation of radiotherapy treatment with optimal transport using a concept for transportation of Ricci-flat time-varying dose–volume}

\author{Yusuke Anetai\orcidlink{0000-0002-2284-7582}{$^{1,*}$} , Jun'ichi Kotoku{$^2$} }
 \affiliation{
 {$^{1}$Department of Radiology, Kansai Medical University, Shin-machi 2-5-1, Hirakata-shi, Osaka 573-0101, Japan.} \\
 {$^{2}$Graduate School of Medical Care and Technology, Teikyo University, 2-11-1 Kaga, Itabashi-ku, Tokyo, Japan.}
 }

\begin{abstract}
In radiotherapy, the dose–volume histogram (DVH) curve is an important means of evaluating the clinical feasibility of tumor control and side effects in normal organs against actual treatment. Fractionation, distributing the amounts of irradiation, is used to enhance the treatment effectiveness of tumor control and mitigation of normal tissue damage. Therefore, dose and volume receive time-varying effects per fractional treatment event. However, the difficulty of DVH superimposition of different situations prevents evaluation of the total DVH despite different shapes and receiving dose distributions of organs in each fraction. However, an actual evaluation is determined traditionally by the initial treatment plan because of summation difficulty. Mathematically, this difficulty can be regarded as a kind of optimal transport of DVH. For this study, we introduced DVH transportation on the curvilinear orthogonal space with respect to arbitrary time ($ T $), time-varying dose ($ D $), and time-varying volume ($ V $), which was designated as the TDV space embedded in the Riemannian manifold. Transportation in the TDV space should satisfy the following: (a) the metrics between dose and volume must be equivalent for any fractions and (b) the cumulative characteristic of DVH must hold irrespective of the lapse of time. With consideration of the Ricci-flat condition for the $ D $-direction and $ V $-direction, we obtained the probability density distribution, which is described by Poisson’s equation with radial diffusion process toward $ T $. This geometrical requirement and transportation equation rigorously provided the feasible total DVH.
\end{abstract}


\maketitle

\section{\label{sec:level1} Introduction \protect\\ }

	Even well-known quantities do not easily add up due to different circumstances. For instance, when we have two pens, “two” is a countable quantity, but the amount of ink of these pens is not necessarily “two”. Because a pen has a variety of shape and capacity with respect to its refill, evaluated as “three” milliliters could be rationale under certain circumstances. The same logic can be applied to “proportions”. We have difficulty of the summation of the proportion because of such quantity 0.3 or 0.5 based on the non-uniform quantity allowing diverse conditions. In actual, this type of summation problem accounting different situations existing has remained unsolved for a long time in radiotherapy.
	
	Radiotherapy delivers a radiation dose to a patient with day-fractionated treatment. A tumor has higher dose response than normal tissues, which leads to differences in repair and damage between the tumor and normal tissues \cite{ref01,ref02}. Therefore, radiotherapy uses fractionated irradiation, which assumes the prescribed dose and dose distribution are equally divided, to increase the likelihood of tumor suppression and to decrease the likelihood of early and late effects on normal tissues \cite{ref03}. The complex dose distribution is implemented under patient-specific organ conditions (Fig.1a–b). As for the treatment evaluation, dose–volume histogram (DVH) provides information related to organ volume in relation to the received radiation dose (Fig. 1c–d), which can compactly indicate a complex dose distribution in an organ with dose–volume indices (DVIs), $ V_x $ (the organ volume receiving dose $ x $) or $ D_y $ (the dose given for organ volume $ y $) \cite{ref04,ref05,ref06}. Consequently, DVH is used widely in the radiotherapy community to estimate both therapeutic feasibility and clinical propriety for tumor control or normal organ side effects. Almost all reported radiotherapeutic effects or clinical outcomes are based on the DVIs of DVH or a similar indices \cite{ref04,ref06}. From an oncologist’s perspective, the DVH and DVI are important tools for ready assessment of clinical effects related to dose distribution. 

	Nevertheless, regarding fractionated irradiation, several patient-specific organ dose–volume conditions exist, such as size changes, deformation of shape, unexpected intensity modulation because of involuntary motions, and patient setup uncertainties. \cite{ref07} Different dose–volumes for different situations can be evaluated for each treatment (Fig. 1e). The total dose to an organ object volume is key to accurately evaluating treatment or side effects. However, summation difficulties arise in total dose–volume estimation due to the diverse conditions of the organ with respect to dose, which we define as the DVH summation problem. As a consequence, the dose-volume uncertainty becomes one of the factors leading to a wide range in the estimation of radiobiological damage for normal tissues.

	Many studies have assessed DVH summation from a statistical perspective, investigating aspects such as patient-setup and margins \cite{ref08,ref09,ref10}, treatment machine errors \cite{ref11, ref12}, dose calculation errors \cite{ref13, ref14, ref15}, temporal or spatial errors \cite{ref16, ref17, ref18}, deformable image registration (DIR) \cite{ref19, ref20, ref21, ref22, ref23}, and the direct uncertainties of DVH \cite{ref24, ref25, ref26, ref27, ref28}. Nevertheless, in each case, only multiplication of the fractionation or specific superimpositions toward the treatment planning is considered, which does not account comprehensively for all different conditions \cite{ref29}. Traditional logic can only account patient condition that limited to the initial treatment planning. However, considering adaptive radiotherapy that has become a new treatment style accounting more reflective of the patient’s condition on the day \cite{ref30, ref31, ref32}, the summarized evaluation of different clinical conditions per treatment day on the patient should be resolved. Our strategy provides the same quantity that encompass a variety of patient's organ conditions. Using this quantity, we aimed to obtain the summation considering different situations.
	
	The planned DVH to the actual fractionated DVH can be regarded as optimal transport. The optimal transport theory has been studied as a Monge–Kantorovich problem, which leads to the interpretation of minimization of Riemannian distance defined by the Wasserstein distance. Non-negative Ricci curvature plays an important role in this theory as a curvature-dimension condition and yields Gromov–Hausdorff convergence of the measure space \cite{ref33}. For this study, we applied this fact to the total DVH convergence. Time-varying DVH represents that dose   and volume   of every fraction have a different meaning compared to that when the plan was created. A prescribed dose of 1 Gy for 1 $\mathrm{cm}^3 $ on the same organ including variable situations because of the reasons described above. Therefore $D=1$  to $V=1$ are time-varying. However, some creative efforts to bring the treatment closer to the plan such as image-guided radiotherapy (IGRT) have been introduced into each treatment. This situation can be interpreted as follows: the basis vector of $D$, $\mathbf{e}_D$, and the basis vector $V$, $\mathbf{e}_V$, are time-varying from the time of planning to the actual treated, but the random fluctuation is not so strong. Nevertheless, every fractionated treatment event can be assessed by a Euclidean $D$ and $V$ relation. From a different perspective, for this problem, it is natural to assume optimal transport between the plan and every treatment event based on Ricci-flat space \cite{ref34}, which allows numerical calculation of the transport in Euclidean conditions. Herein, we propose a newly developed method for dose–volume transportation in the specific time-varying dose–volume (TDV) space to estimate the feasible total DVH or DVI.

\begin{figure}[b]

\includegraphics[width=0.95\columnwidth]{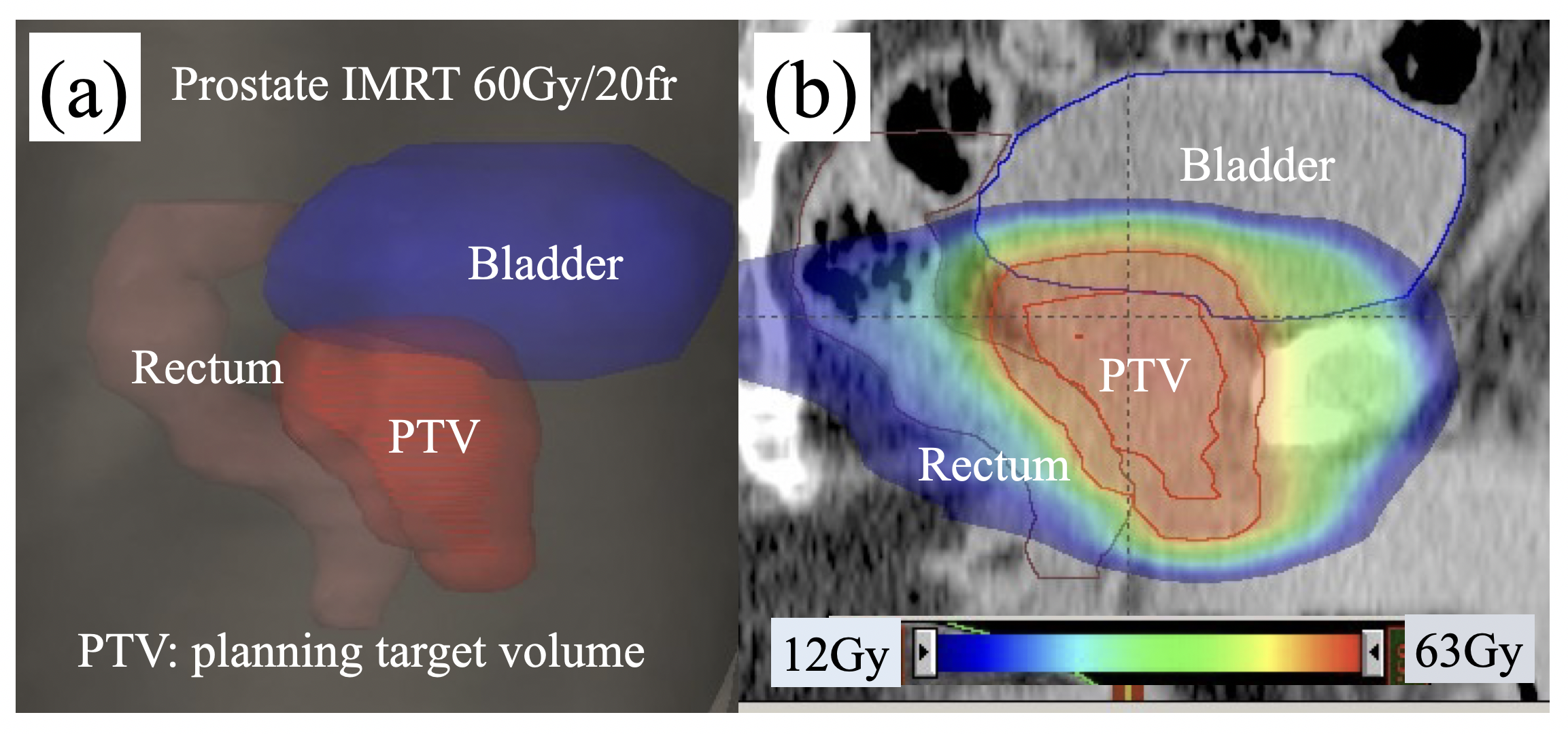}
\includegraphics[width=0.95\columnwidth]{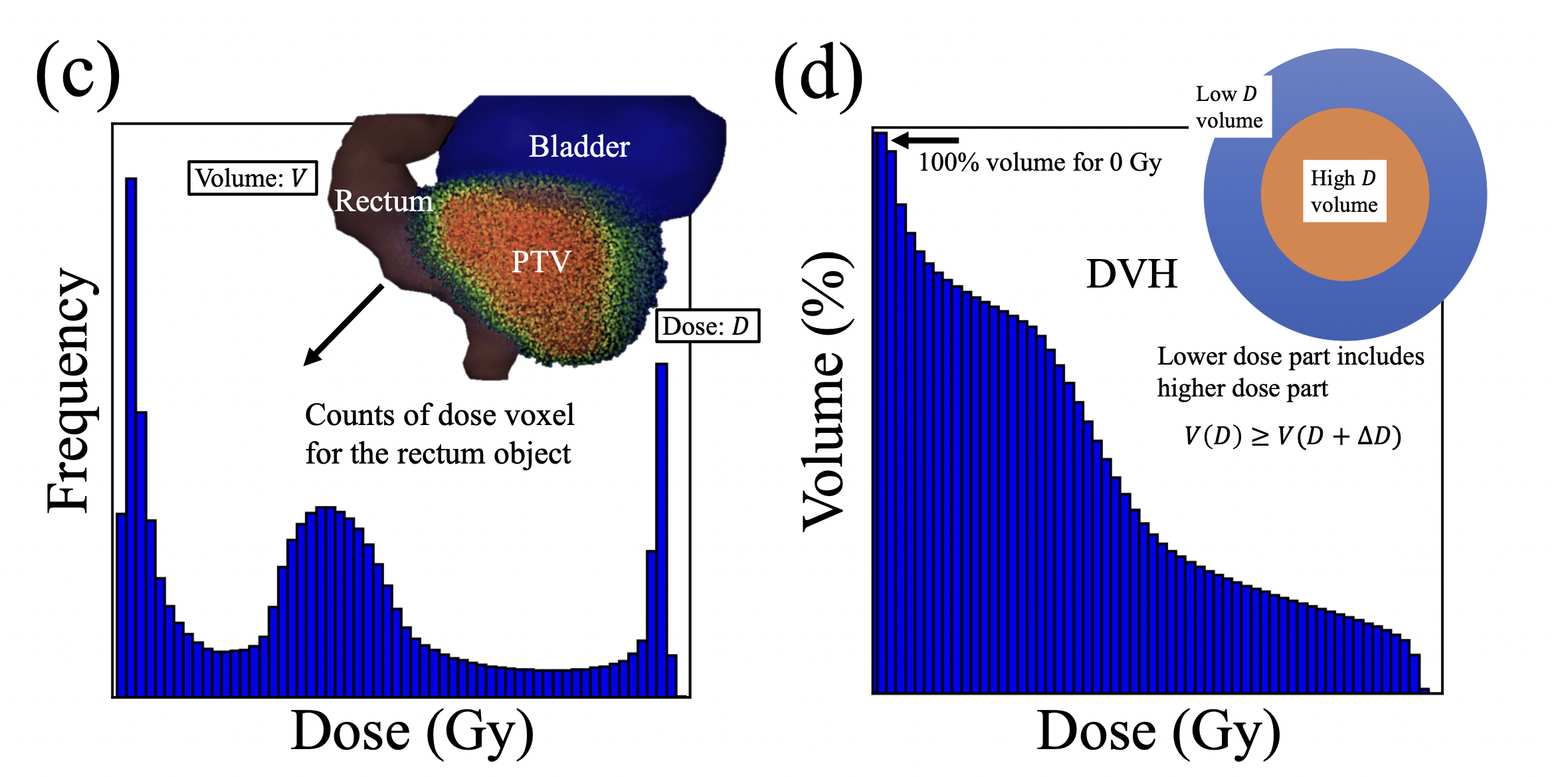}
\includegraphics[width=0.95\columnwidth]{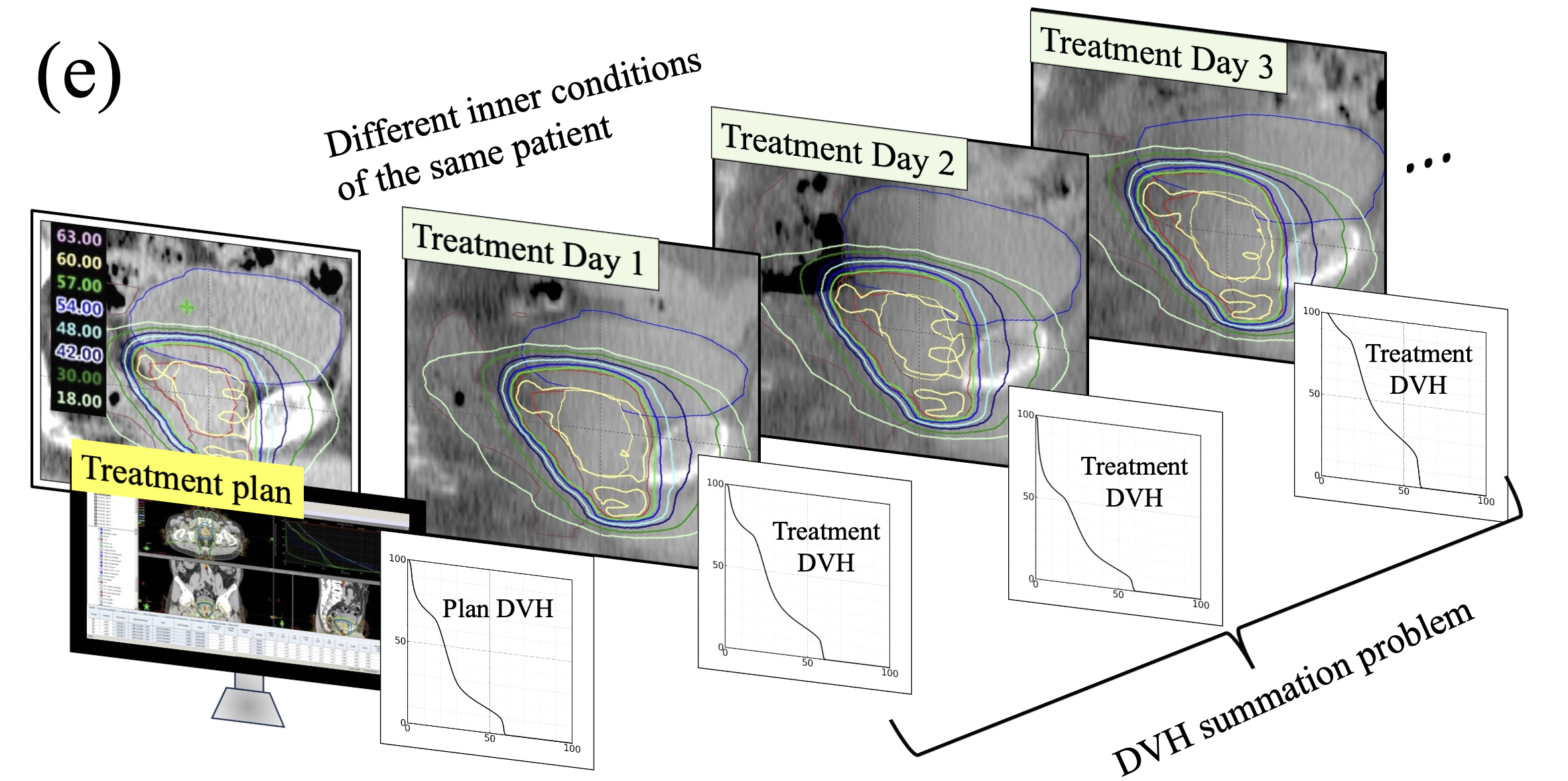}
\caption{\label{fig:epsart} 
Radiotherapy treatment planning for prostate using intensity modulated radiation therapy (IMRT), which enables delivery of a high dose to tumor/cancer lesions and delivery of a low dose to normal tissue regions. A tumor target (prostate) is margined as a planning target volume (PTV) to prescribe a specific radiation dose (60 Gy) by 20-fractionated treatment. (a) PTV-bladder-rectum locational relations. (b) Dose distribution and related organ volume. (c) In the planned dose distribution, dose deposition can be counted as every voxel per organ object volume. (d) Dose–volume histogram (DVH) for the rectum. (e) Fractionated treatment allows a large variety of different organ states for the same targeted organ. Consequently, DVH fluctuation caused by different dose distribution occurs, creating a DVH summation problem.}

\end{figure}
	
\section{\label{sec:level1} Methods \protect}

\subsection{\label{sec:level2} Time-Dose-Volume (TDV) space}
DVH curve transportation is characterized by the Time–Dose–Volume (TDV) manifold $ \cal{M}_{TDV} \in \mathbb{R_{+}} $, which denotes the topological space composed of time-varying dose $ D  \in \mathbb{R_{+}}  $  time-varying volume $ V  \in \mathbb{R_{+}}  $ , and time $ T   \in \mathbb{R_{+}} $ , including any $ D $, $ V $, and $ T $ data. To assess the TDV manifold, local coordinate maps are defined by a tangential space $ \partial  \cal{M}_{TDV} $. 

	One can let $ D $  and $ V $ relations over a long time $ T $ be the tangent vector space $ \partial  \cal{M}_{TDV} $ , designated as TDV space. Additionally, $ \partial  \cal{M}_{TDV} $ is set as the basis vector set for local coordinate map projection on  $ \cal{M}_{TDV} $. 
Here, $ T \geq 0 $, $ D \geq 0 $, and $ V \geq 0 $ are constraints for the independent variables. One can set TDV space in a Riemannian manifold, which is a submanifold of $ \cal{M}_{TDV} $. The boundary between the surface maps is connected by a Levi–Civita connection, with connection coefficients defined as $ \Gamma^{k}_{ij} $ with respect to coordinate suffixes $ i, j, k$. The TDV spatial metric is defined using a Riemannian metric tensor $ \mathrm{g}_{ij} $. Here, we define TDV space as the following orthogonal geometric system:

\begin{equation}
ds^2 = dT^2 + {\xi}^2(T)dD^2 + {\eta}^2(T)dV^2
\label{eq:one},
\end{equation}

where  $ \xi $ and $ \eta $ are metric parameters for time-varying distortion of space, and where $ ds $ is the line element of local coordinates. This coordinate system is convenient in cases with the same dose quantity $ D $ and volume quantity $ V $ representing different situations from time to time. The metric tensor $ \mathrm{g}_{ij} $ for the TDV space is therefore
\begin{eqnarray}
	\mathrm{g}_{ij} = 
		\left(
		\begin{matrix}
			1 & 0 & 0 \\ 0 & {\xi}^2 & 0 \\ 0 & 0 &{\eta}^2 
		\end{matrix}
		\right).
\end{eqnarray}

Local coordinates are connected using the affine connection coefficient  $ \Gamma^{k}_{ij} $ as
\begin{equation}
\Gamma^{k}_{ij} = \frac{1}{2} \mathrm{g}^{kl} 
	\left(
		\partial_{i} \mathrm{g}_{lj} +  \partial_{j} \mathrm{g}_{li} -\partial_{k} \mathrm{g}_{ij}
	\right),
\end{equation}
where $ l $ also represents a coordinate suffix. In this article, for some  tensor $ X^{ij} $ and basis vector $ \mathrm{e}_{ij} $ set, the following Einstein notation is adopted.
\begin{equation}
X^{ij} \mathrm{e}_{ij} = \sum_{i} \sum_{j} X^{ij} \mathrm{e}_{ij}.
\end{equation}

Then, Levi–Civita connection coefficients $ \Gamma^{k}_{ij} $ are obtained as (Appendix A)
\begin{equation}
	\begin{split}
		\Gamma^{T}_{DD} &= -\xi \frac{\partial \xi}{\partial T} = -\xi \dot{\xi}, \\
 		\Gamma^{T}_{VV} &= -\eta \frac{\partial \eta}{\partial T} = -\eta \dot{\eta}, \\
 		\Gamma^{D}_{TD} &=  \Gamma^{D}_{DT} = \frac{\dot{\xi}}{\xi}, \\
		\Gamma^{V}_{TV} &=  \Gamma^{V}_{VT} = \frac{\dot{\eta}}{\eta}, \\
 	\end{split}
\end{equation}

\begin{figure}[b]

\includegraphics[width=0.95\columnwidth]{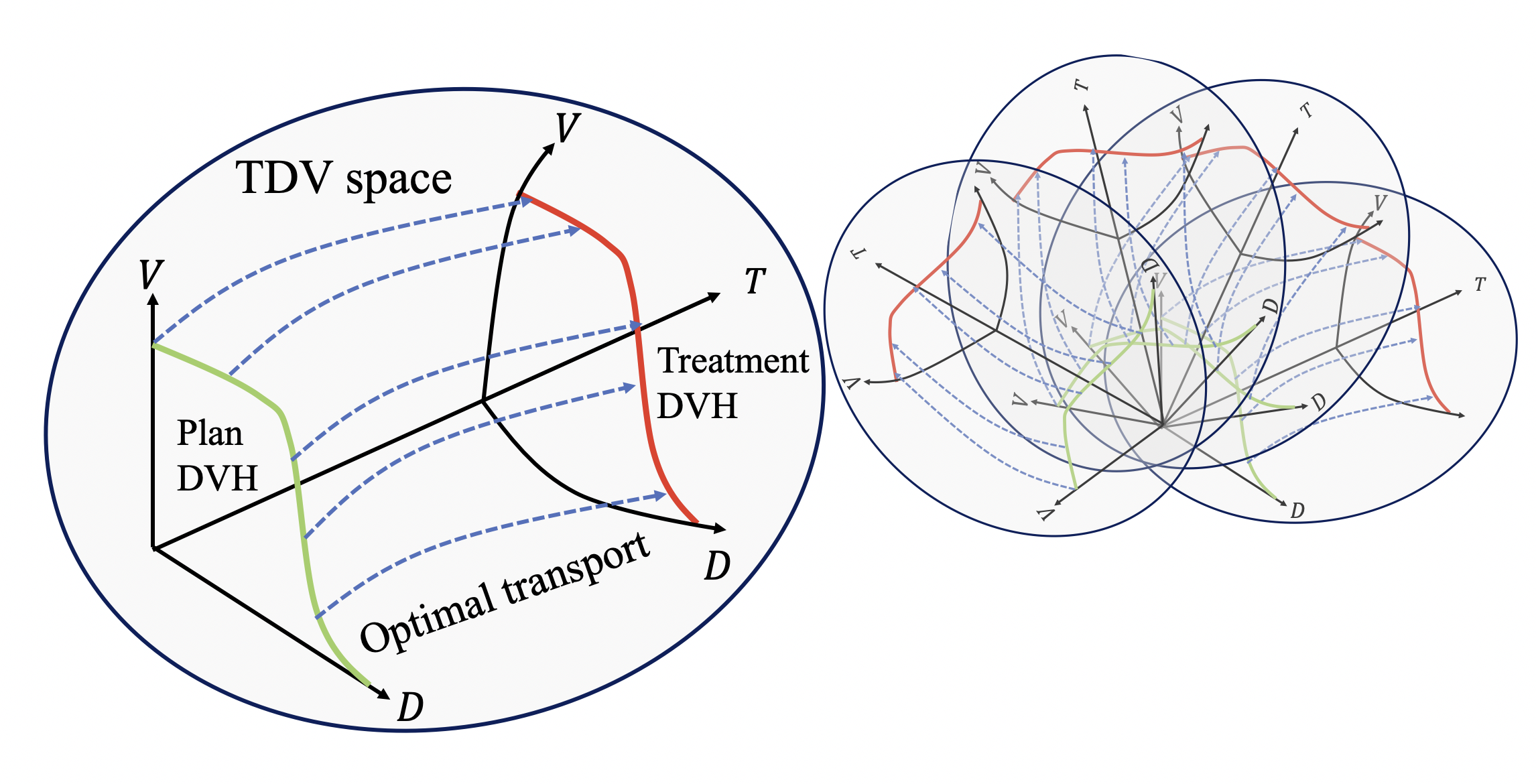}
\caption{\label{fig:epsart} 
Multiple Time–Dose–Volume (TDV) spaces are superimposed on a Riemannian manifold. Optimal transport is conducted along with the geodesics. The probabilistic characteristic of optimal transport can allow superimposition of the different situations of DVH.}

\end{figure}

\subsection{\label{sec:level2} DVH transportation with optimal transport in TDV space}

Let DVH transportation be a smooth transportation (continuous deformation) from the first planned DVH to the actual treated state DVH in TDV space with respect to $ T $ , where $ T $ is an arbitrary time interval, and where $ D $ and $ V $ respectively represent the dose and volume. Also, $ T $ does not necessarily correspond to actual time passing. The several conditions below should be applied to DVH transportation. $\\$

\begin{flushleft} \textbf{Condition 1}: Descending rule of  $ V(D) $  \end{flushleft}
From constraint of DVH characteristic, the following descending rule with respect to $ V $ holds for any time $ T $ with respect to increments of dose $ \Delta D \geq 0 $  (Fig. 1d), as
\begin{equation}
	V(D) \geq V(D+\Delta D).
\end{equation}

\begin{flushleft} \textbf{Condition 2}: Partially Ricci-flat condition, equivalence of $ D $ and $ V $\end{flushleft}
The relation  $ D $ and $ V $ is equivalent to Euclidean space for any time $ T $. According to \textbf{Condition 1}, the area metric remains Euclidean. Therefore, Ricci curvature $ \mathrm{Ric}(\cdot, \cdot)$  with respect to the TDV space is
\begin{equation}
	\mathrm{Ric}(D, D) = \mathrm{Ric}(V, V) = \mathrm{Ric}(D, V) = \mathrm{Ric}(V, D) = 0
\end{equation}

Here, we introduce Ricci curvature tensor 
\[
	R_{ij} = \partial_{m}  \Gamma^{m}_{ij} - \partial_{j}  \Gamma^{m}_{im} 
				+ \Gamma^{m}_{nm} \Gamma^{n}_{ij} - \Gamma^{m}_{nj} \Gamma^{n}_{im} 
\]
where $ R_{ij} $ represents the mean sectional curvature for the directions of the TDV space, where $ i, j, m, n $  are contra-variant or co-variant suffixes of the tensor. This curvature represents a degree of deviation of the sphere in a TDV space from the Euclidean space in terms of the volume metric. The following expressions are applicable (Appendix B):
\begin{equation}
	\begin{split}
	R_{TT} &= -\partial_{T} \left( \Gamma^{D}_{TD} + \Gamma^{V}_{TV} \right)
					-\left( \Gamma^{D}_{DT} \Gamma^{D}_{TD} - \Gamma^{V}_{VT} \Gamma^{V}_{TV} \right), \\ 
	R_{DD} &= \partial_{T}  \Gamma^{T}_{DD} + \left(  \Gamma^{V}_{TV} -  \Gamma^{D}_{DT} \right)  \Gamma^{T}_{DD}, \\
	R_{VV} &= \partial_{T}  \Gamma^{T}_{VV} + \left(  \Gamma^{D}_{TD} -  \Gamma^{V}_{VT} \right)  \Gamma^{T}_{VV}, \\
	R_{TD} &= R_{DT} = R_{TV} = R_{VT} = R_{DV} = R_{VD} = 0.   
	\end{split}
\end{equation}

These relations eliminate the Cotton tensor, that is conformal invariance, retaining the orthogonal system. To satisfy the equivalence of $D$ and $V$, $ R_{DD} = 0 $ and $ R_{VV} = 0 $ are required. Then, $ \xi \eta = aT+b $ is obtained. Hence, 
\begin{equation}
	\begin{split}
		\xi &= (aT+b)^{\lambda}, \\
		\eta &= (aT+b)^{1-\lambda},
	\end{split}
\end{equation}
where $ \lambda \in \mathbb{R} $ is a coordinate factor $ a(D,V) $ and $ b(D,V) $ are time-independent coefficients. When $ T=0 $(plan), the TDV space is a completely Euclidean coordinate. For that reason, $ \xi(0) = 1 $ and $ \eta(0) = 1 $. Therefore,  $ b(D,V) = 1 $. Moreover, the equivalence of $ D $ and $ V $ determines $ \lambda = 1/2 $. Consequently, \textbf{Condition 2} can be rewritten as a simple coordinate condition, as
\begin{equation}
	\xi = \eta = \sqrt{aT+1},
\end{equation}
Since $\xi$ and $\eta$ are time-dependent and not a function of $D$ and $V$, we fixed $a$ as a constant. 

\begin{flushleft} \textbf{Condition 3}: Optimal transport \end{flushleft}
DVH transportation should be optimal transport in the TDV space. Now we have elements from sets of two probability distributions $ \mathrm{x} \in \mathrm{X} $ and $ \mathrm{y} \in \mathrm{Y} $. Let the $ p $-Wasserstein distance $ W_p $ be
\begin{equation}
	W_p(\mu, \nu) = \left( \inf_{\pi \in \Pi(\mu,\nu)} 
	\int_{\cal{M} \times \cal{M}} \mathrm{d}( \mathrm{x}, \mathrm{y}) d\pi( \mathrm{x}, \mathrm{y})  \right)^{\frac{1}{p}}
\end{equation}
where $ p \in [1,\infty] $ and probability measures $ \mu, \nu $ on metric space $ \cal{M} $ and distance function $ \mathrm{d}(\cdot,\cdot) $ satisfy joint probability $ \pi (\mu, \nu) $, which is included by coupling set $ \Pi(\mu, \nu)$  \cite{ref35}.

In general, optimal transport in the Riemannian surface requires $ p=2 $, and minimization of the cost \cite{ref36, ref37, ref38}. When the Ricci curvature is a non-negative condition, the Wasserstein metric propagates geodesic convexity, which engenders the convergence of probability measures \cite{ref33, ref39, ref40}. With application of these facts to the DVH transportation in the TDV space, we can obtain the total estimated DVH via coupling set $ \Pi(\mu, \nu)$ between the plan and every actual treatment because probability $ \pi (\mu, \nu) $ can be superimposed. We can analogically redefine $ D $  and $ V $ respectively as the relative dose and relative organ volume such as the dose per fractionated planned dose and organ volume per estimated volume at the treatment, which satisfy $ 0 \leq D \leq 1 $ and $ 0 \leq V \leq 1 $. Figure 2 depicts the scheme for the total DVH estimation considering multiple TDV spaces. $\\$

\begin{flushleft} \textbf{Condition 4}:  Irrotational flow with no divergence and the transportation equation \end{flushleft}
Optimal transport is performed according to geodesics satisfying irrotational and no divergence in the Riemannian manifold. The flow satisfies no inversed time and no stationary point (conservative vector field). Let the velocity vector field in TDV space be $A^l $ (where $l$ denotes a coordinate suffix), described as
\begin{equation}
	A^l \partial_{l} = \left( \frac{\partial \Phi}{\partial T} \right) \partial_{T} 
				+  \frac{1}{{\xi}^2} \left( \frac{\partial \Phi}{\partial D} \right) \partial_{D}
				+  \frac{1}{{\eta}^2} \left( \frac{\partial \Phi}{\partial V} \right) \partial_{V},
\end{equation}
where $ \partial_{l} $ represents the basis vector, and where $ \Phi := \Phi(T,D,V) $  is a scalar potential. Then,  $ A^l $ satisfies
\begin{equation}
	\begin{split}
		& \mathrm{rot}(A^l) = \textbf{0}, \\ &\therefore  A^l = \mathrm{grad}(\Phi) = \partial_{ij} \Phi \mathrm{g}^{ij},
	\end{split}
\end{equation}
and
\begin{equation}
	\begin{split}
		& \mathrm{div}(A^l) = 0, \\ &\therefore  \partial_{l} A^l + \Gamma^{l}_{nl} A^n = 0.
	\end{split}
\end{equation}
From Equations (5) and (12)–(14), the following equation in the TDV space is required for the flow:
\begin{equation}
	\frac{ {\partial}^2 {\Phi}}{\partial T^2}
	+ \left(  \frac{\dot{\xi}}{\xi} +  \frac{\dot{\eta}}{\eta} \right) \frac{ \partial {\Phi} }{\partial T}
	+ \frac{1}{{\xi}^2}  \frac{ {\partial}^2 {\Phi} }{\partial D^2} 
	+ \frac{1}{{\eta}^2} \frac{ {\partial}^2 {\Phi} }{\partial V^2} = 0.
\end{equation}

Now, the restriction of Ricci curvature can prevent unnecessary divergence of the space (curvature-dimension condition); then, the TDV space can converge. From \textbf{Condition 2} and $R_{TT}$ term of Eq. (8) , Ricci tensor  $R_{ij} $ of the TDV space can be contracted to the $ R_{TT} $ term (because other terms are 0), hence,
\begin{equation}
	R_{TT} = \frac{2a^2\lambda(1-\lambda)}{(aT+1)^2} = \frac{a^2}{2(aT+1)^2} \geq 0.
\end{equation}

Consequently, transportation via Eq. (15) satisfies optimal transport. Specifically, Eq. (15) can be rewritten as
\begin{equation}
	\begin{split}
		\frac{1}{aT+1} & 
		\left( \frac{ {\partial}^2 {\Phi} }{\partial T^2} + \frac{ {\partial}^2 {\Phi} }{\partial D^2} + \frac{ {\partial}^2 {\Phi} }{\partial V^2} \right) \\ 
		&+\frac{aT}{aT+1} \left(  \frac{ {\partial}^2 {\Phi}}{\partial T^2} + \frac{1}{T} \frac{\partial \Phi}{\partial T} \right)= 0,
	\end{split}
\end{equation}
where the first term is Poisson’s equation and where the second term represents the cylindrical diffusive process. Therefore, 
\begin{equation}
	\begin{split}
	&  \frac{ {\partial}^2 {\Phi} }{\partial T^2} + \frac{ {\partial}^2 {\Phi} }{\partial D^2} + \frac{ {\partial}^2 {\Phi} }{\partial V^2} = -K, \\
	&  \frac{ {\partial}^2 {\Phi}}{\partial T^2} + \frac{1}{T} \frac{\partial \Phi}{\partial T} = \frac{K}{aT},
	\end{split}
\end{equation}
are required simultaneously, where $ K:= K(D,V) \geq 0 $ . The second term is solved as
\begin{equation}
	\Phi(T,D,V) = C_{1}(D,V) + C_{2}(D,V) \mathrm{ln} T + K(D,V)\frac{T}{a},
\end{equation}
where $C_1$ and $C_2$ are some functions of $ D $ and $ V $. Then, We can obtain the relation
\begin{equation}
	\frac{\partial \Phi}{\partial T} =  \frac{C_{2}(D,V)}{T} + \frac{K(D,V)}{a},
\end{equation}
which relates to the dose–volume variance reduction according to $1/T$ with an increase of the fraction reported by Unkelbach \cite{ref41, ref42}.

\begin{flushleft} \textbf{Condition 5}:  Diffusion coefficient $ 1/a $ \end{flushleft}
Now, parameter $ a $ is a matter to solve the transportation equation numerically. If $ a = 0 $ , then the solution of Eq. (17) is the steady-state according to Laplace’s equation in the TDV space. The second equation of Eq. (18) suggests radial diffusion $ (a < 0) $ or unnatural condensation $(a > 0)$ process toward the  $T$-direction according to diffusion coefficient $1/a$. Term $ (K/aT) $ represents a scaled source term of the diffusion/condensation of the system with an increase of $T$. The condition of $ a < 0 $ is appropriate because of the positive diffusion coefficient. Moreover, if $a < 0$ is set, then the metric parameters $\xi=\eta=\sqrt{aT+1}$ can be expected to satisfy $|a|T \leq 1$  because of $T \geq 0$. Therefore, the following relation is required.
\begin{equation}
	0 \leq T \leq  \frac{1}{|a|}, a<0.
\end{equation}

Although $ T $ is defined as an arbitrary time parameter satisfying Eq. (21), $ T $ can be redefined as the relative elapsed time from the plan to the treatment in the duration $ [0,1] $ with a scale factor $t_0 = 1/|a|$, therefore $ 0 \leq T/ t_{0} \leq  1$. The key parameters $T$, $D$, and $V$ are consistent in the TDV space as the probabilistic feature of optimal transport.

\subsection{\label{sec:level2} Numerical calculation and its specific parameters}

We encountered the difficulty that Equations (18) and (19) introduce three undetermined parameters $C_1$, $C_2$ and $ K $ to numerically solve the transport calculation. Then, we directly calculated Eq. (17), which is rewritten in the final form for numerical analysis using combined Leap-frog and Crank–Nicolson methods as
\begin{equation}
	\begin{split}
		&(aT+1)	
		\frac{\Phi^{\mathrm{k+1}}_{\mathrm{i,j}}-2\Phi^{\mathrm{k}}_{\mathrm{i,j}} + \Phi^{\mathrm{k-1}}_{\mathrm{i,j}}}{(\Delta T)^2} 
		+ a\frac{\Phi^{\mathrm{k+1}}_{\mathrm{i,j}}-\Phi^{\mathrm{k}}_{\mathrm{i,j}}}{(\Delta T)} \\
		&+ \frac{1}{2}
		\left[  
		\frac{\Phi^{\mathrm{k+1}}_{\mathrm{i,j+1}}-2\Phi^{\mathrm{k+1}}_{\mathrm{i,j}} + \Phi^{\mathrm{k+1}}_{\mathrm{i,j-1}}}{(\Delta D)^2}
		+
		\frac{\Phi^{\mathrm{k}}_{\mathrm{i,j+1}}-2\Phi^{\mathrm{k}}_{\mathrm{i,j}} + \Phi^{\mathrm{k}}_{\mathrm{i,j-1}}}{(\Delta D)^2}  
		\right] \\
		&+ \frac{1}{2}
		\left[  
		\frac{\Phi^{\mathrm{k+1}}_{\mathrm{i+1,j}}-2\Phi^{\mathrm{k+1}}_{\mathrm{i,j}} + \Phi^{\mathrm{k+1}}_{\mathrm{i-1,j}}}{(\Delta V)^2}
		+
		\frac{\Phi^{\mathrm{k}}_{\mathrm{i+1,j}}-2\Phi^{\mathrm{k}}_{\mathrm{i,j}} + \Phi^{\mathrm{k}}_{\mathrm{i-1,j}}}{(\Delta V)^2}  
		\right] \\
		& = 0,
	\end{split}
\end{equation}
where $ \mathrm{i,j,k} $ are defined respectively as increment indices for row, column, and layer directions corresponded to the simulation coordinate of $V$, $D$, and $T$ directions. Also, $ \Delta D$, $ \Delta V$, and $ \Delta T$ are increment parameters. Time lapse $T$ is followed by $ T = t_0\mathrm{k}\Delta T$.  As a result, we can implicitly calculate Eq. (17) as
\begin{equation}
	\begin{split}
		\Phi^{\mathrm{k+1}}_{\mathrm{i,j}} = 
			&\frac{2(a t_0 \mathrm{k} \Delta T +1)+a\Delta T + {\beta}^2 + {\gamma}^2}
				{a t_0 \mathrm{k} \Delta T +1+a\Delta T - {\beta}^2 - {\gamma}^2}
				\Phi^{\mathrm{k}}_{\mathrm{i,j}} 
			\\
			&- \frac{a t_0 \mathrm{k} \Delta T +1}
				{a t_0 \mathrm{k} \Delta T +1+a\Delta T - {\beta}^2 - {\gamma}^2}
				\Phi^{\mathrm{k-1}}_{\mathrm{i,j}} 
			\\
			&- \frac{
			{\beta}^2 
			\left( \Phi^{\mathrm{k+1}}_{\mathrm{i+1,j}} + \Phi^{\mathrm{k+1}}_{\mathrm{i-1,j}}
			+ \Phi^{\mathrm{k}}_{\mathrm{i+1,j}} + \Phi^{\mathrm{k}}_{\mathrm{i-1,j}} \right) 
			}
			{2(a t_0 \mathrm{k} \Delta T +1+a\Delta T - {\beta}^2 - {\gamma}^2)}
			\\
			&- \frac{
			{\gamma}^2 
			\left( \Phi^{\mathrm{k+1}}_{\mathrm{i,j+1}} + \Phi^{\mathrm{k+1}}_{\mathrm{i,j-1}}
			+ \Phi^{\mathrm{k}}_{\mathrm{i,j+1}} + \Phi^{\mathrm{k}}_{\mathrm{i,j-1}} \right) 
			}
			{2(a t_0 \mathrm{k} \Delta T +1+a\Delta T - {\beta}^2 - {\gamma}^2)},
	\end{split}
\end{equation}
where $\gamma := \Delta T/ \Delta D$ and $\beta := \Delta T/ \Delta V$.

From Eq. (23), the condition $ a \Delta T + 1 \leq 0$ for $ \mathrm{k} \geq 1$ gives the stable calculation. Considering \textbf{Condition 5}, we set the parameters $\Delta T = 0.1$ ,  $\Delta D = 1.0$, and  $\Delta V = 1.0$ for the numerical calculation satisfying the Courant–Friedrichs–Lewy condition $\Delta T \ll \Delta D, \Delta V$  of computational fluid dynamics. Every treatment receives optimal transport from the initial condition. Then we calculate $\Delta T$ steps until $ t_0\mathrm{k}\Delta T = 1$ to reach the converged event as the total DVH. For that, we set  $ a = -10$, $ t_0 = 0.1$ , $ \mathrm{k} = 10 $. Other detailed calculation conditions are set as presented below.

\begin{flushleft} \textbf{Condition 6}:  Detailed numerical calculation conditions and total DVH \end{flushleft}
We set 1.0 $\%$ per the maximum fractionated dose for  $D$-metric and 1.0 $\%$  $V$-metric for DVH. All DVHs for each obtained treatment event were superimposed as sparse potential point: then, they were scaled to the sum of area to 1 using $101 \times 101$ matrix for the implicit calculation area. The increment indices corresponded to $\mathrm{i} = 0,1, \cdots, 100$, $\mathrm{j} = 0,1, \cdots, 100$, and $\mathrm{k} = 0,1, \cdots, 10$. The Dirichlet boundary condition was applied to the base axes, obtaining the following. 
\[
	\Phi^{\mathrm{k}}_{\mathrm{0,j}} = \Phi^{\mathrm{k}}_{\mathrm{i,0}} = 0, 
	\Phi^{\mathrm{k}}_{\mathrm{100,j}} = \Phi^{\mathrm{k}}_{\mathrm{i,100}} = 0,\]
The initial potential $ \Phi^{\mathrm{-1}}_{\mathrm{i,j}} = 0 $ was applied. When the calculation transfers the next time-step $(\mathrm{k}+1)\Delta T$, we scaled the total sum  $\sum_{i} \sum_{j} \Phi^{\mathrm{k}}_{\mathrm{i,j}} $ to retain the initial total sum $\sum_{i} \sum_{j} \Phi^{\mathrm{0}}_{\mathrm{i,j}} $  for the conservation law. We defined the goal of total DVH as the line plot of the highest potential point (maximized likelihood) in each dose, therefore,
\begin{equation}
	V_{tot}(D_{i}) = \mathrm{max}(\Phi^{\mathrm{k}}_{\mathrm{i,j}} ),
\end{equation}
where $V_{tot}$ is volume of the total DVH with respect to dose. The processing flow to obtain the total DVH is depicted as Fig. 3.

\begin{figure}[b]

\includegraphics[width=0.95\columnwidth]{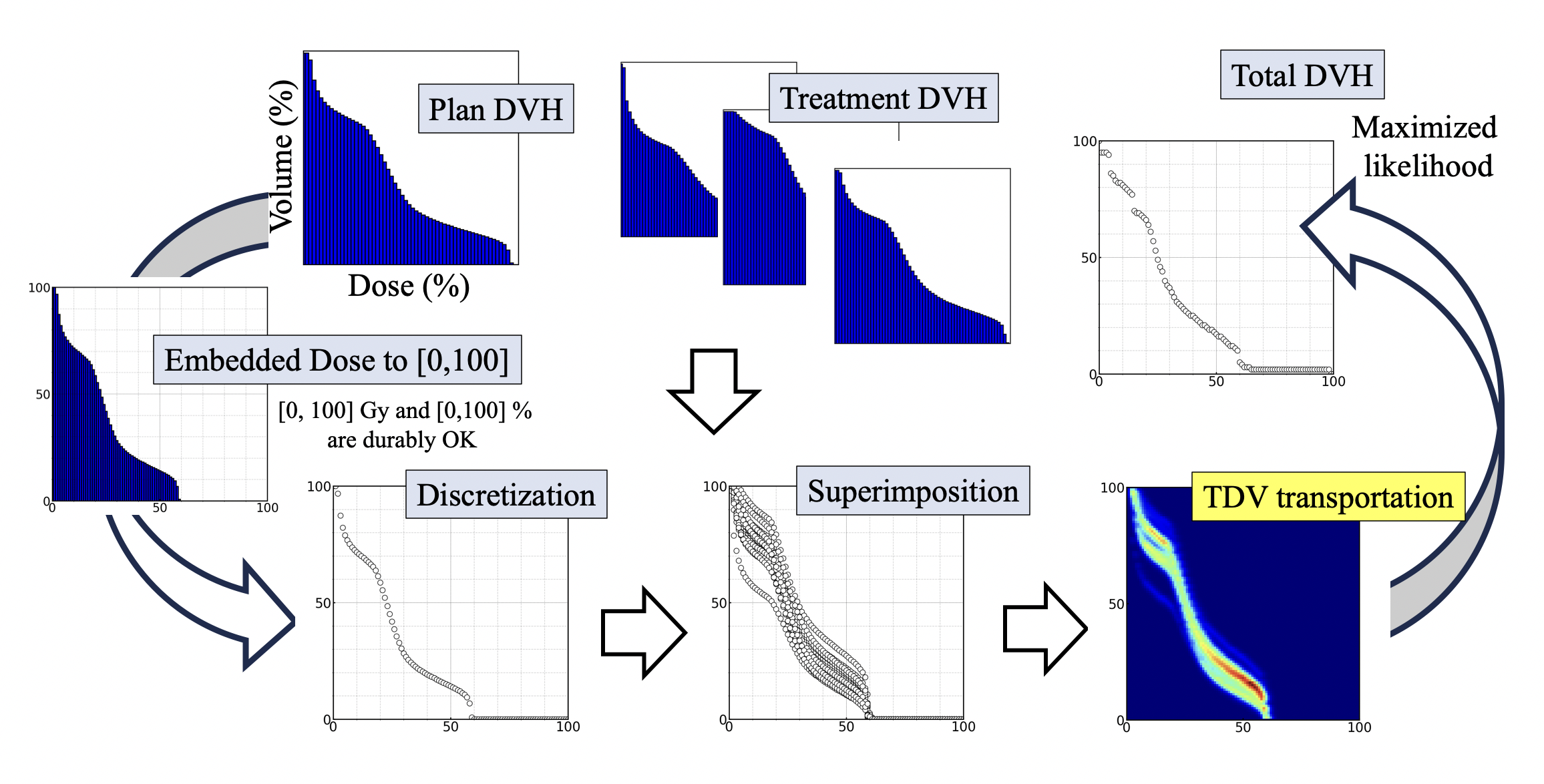}
\caption{\label{fig:epsart}
Processing flow to obtain the total DVH. Plan DVH is obtained from the planned dose distribution. We embedded the dose–volume relation to the durable dose scale $[0, 100]$ Gy or $[0,100]$ \%. To calculate the coupling probability numerically, we discretized the DVH points as a potential distribution with discretized treatment DVHs also. Then we calculate the transportation according to Eq. (23). Total DVH is obtained as the maximum likelihood of the transported potential.}

\end{figure}

\section{\label{sec:level1} Results \protect}

A treatment planning system (Eclipse; Varian Medical Systems Inc., Palo Alto, CA, USA) for clinical use of radiotherapy was used for creation of the IMRT dose distribution, which also calculated DVH for the treatment plan (plan DVH) with the use of computed tomography (CT) and 20-fractionated treatment DVHs using an image-guide cone-beam CT for one prostate patient treated at Kansai Medical University.

\begin{figure}[b]

\includegraphics[width=0.95\columnwidth]{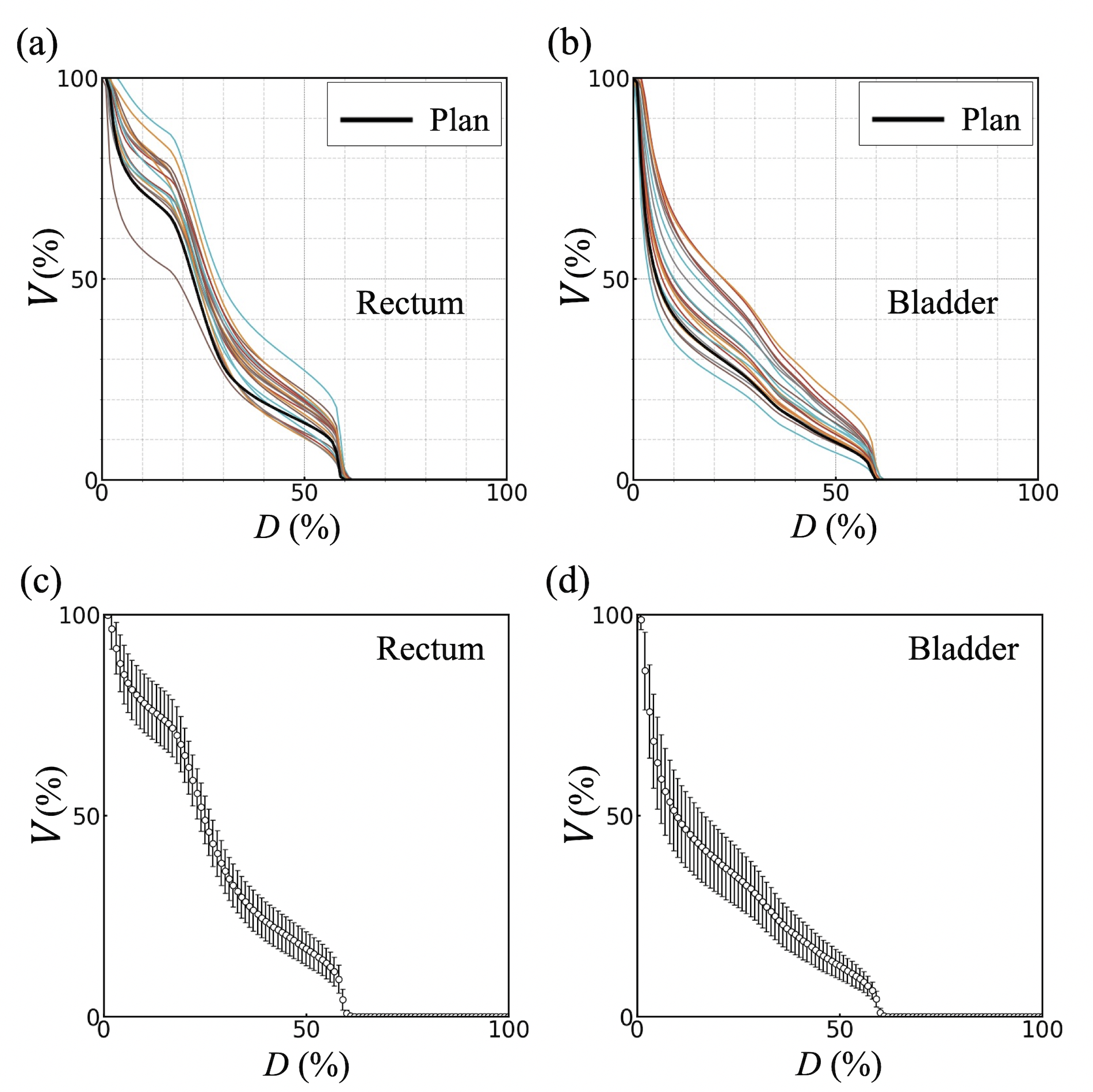}
\caption{\label{fig:epsart} 
(a)–(b) DVHs of the rectum and bladder with respect to plan and treatment. Bold black lines represent the planned DVH. (c)–(d) Average and SD of treatment DVHs of rectum and bladder with respect to 1\%.}

\end{figure}

We specifically examined the rectum and bladder DVHs, which usually have various organ shapes and volumes in each treatment and which have high clinical importance when caring for bleeding. Here, we set 100 Gy = 100 $\%$ as the relative dose for estimating the total DVH. Simultaneously, the fractionated treatment 100$\%$ dose was applied to the equally fractionated plan 100$\%$ dose. DVHs with respect to the rectum and bladder at plan and fractionated treatment in this study are depicted as shown in Figs. 4a–b. According to the variable shape and placements, treatment DVHs for these organs deviated from the plan DVHs. Figures 4c–d portray the average line with standard deviation. Probability density was obtained as TDV transportation of DVHs, as shown in the Figs. 5a–b. 

\begin{figure}[b]

\includegraphics[width=0.95\columnwidth]{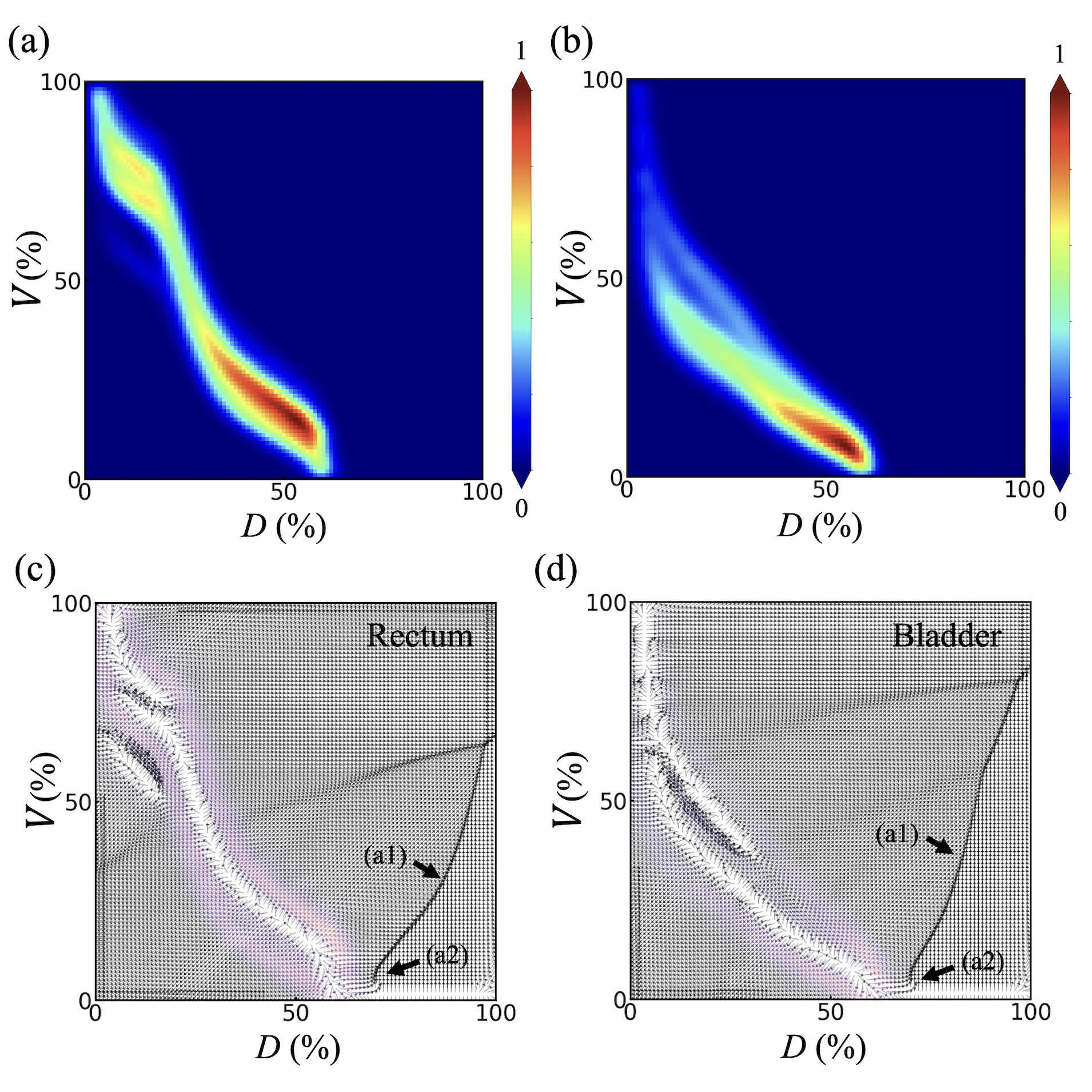}
\caption{\label{fig:epsart} 
TDV transportation against various DVH shapes of Fig. 4(a)–(b). Dose–volume relations represented by probability density via TDV transportation at  $\mathrm{k} = 10 $ with respect to the rectum and bladder. (c)–(d) Flow field of the probability density related to $A^D {\partial}_D + A^V {\partial}_V $ with respect to (a) and (b). Annotation (a1) shows the flow convergence which depicts the confined space of the TDV space; (a2) shows the vector fields outside the TDV space, which is negligible.}

\end{figure}

These results were consistent with the fact that the limited volume of the higher dose is more feasible than the spread volume of the lower dose. Figures 5c–d present the flow field of probability density and satisfaction of irrotational and incompressible flow (\textbf{Condition 2}). Strong flow convergence occurred, which was related to the high sectional curvature point because of the boundary of the TDV space. Consequently, the outside flow of TDV space should be neglected.

\begin{figure}[b]

\includegraphics[width=0.95\columnwidth]{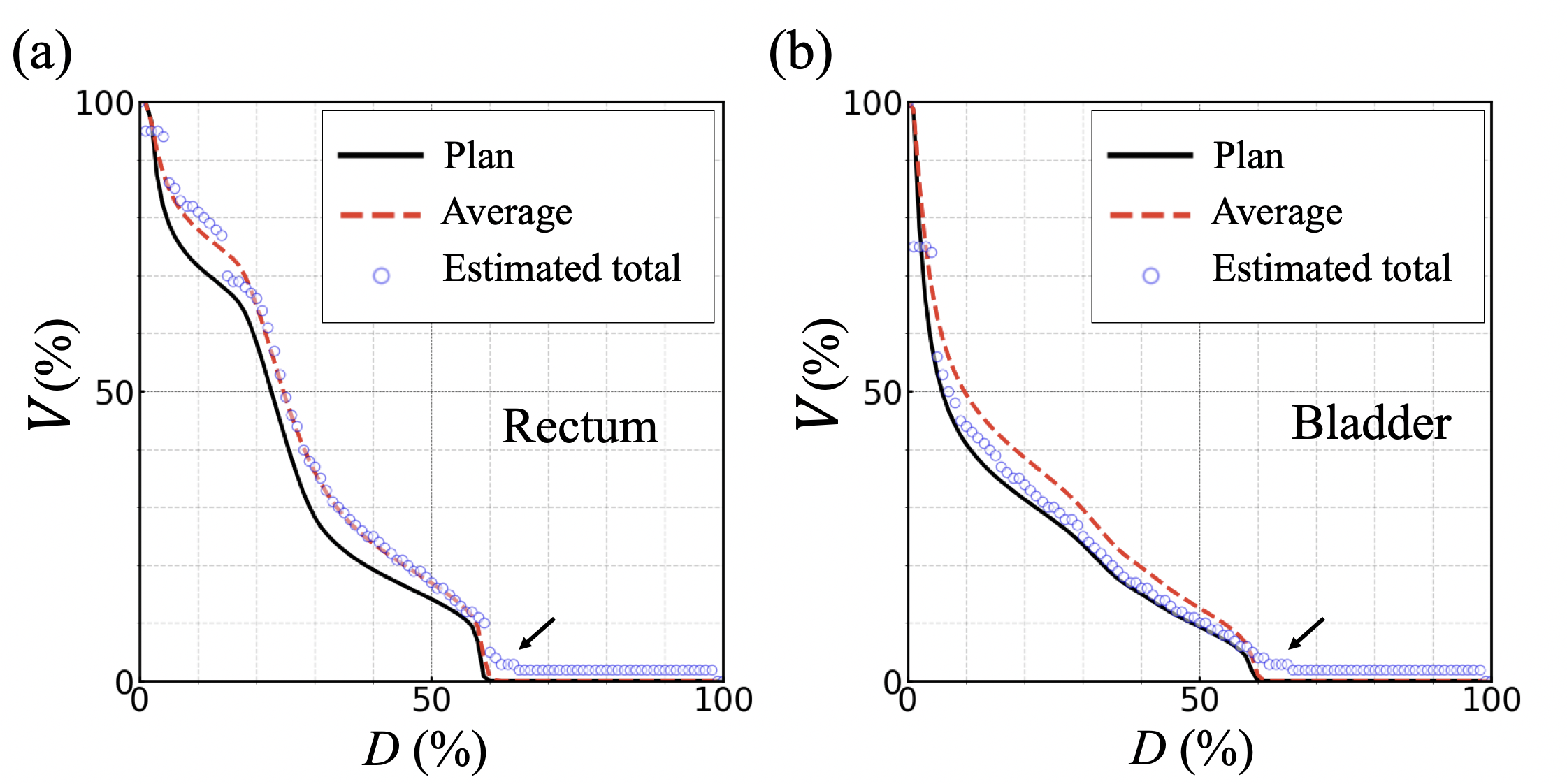}
\caption{\label{fig:epsart} 
Comparisons of the DVHs of plan, average of 20 fractions, and estimation total DVH via TDV transportation at $\mathrm{k} = 10$  with respect to the rectum and bladder. Arrow notations represent inaccuracy of the dose maximum.}

\end{figure}

 Finally, we obtained the total DVH as the most feasible DVH with calculation of the $V$ point that gives the maximum of probability density with respect to $D$ , as depicted in Fig. 6. The total DVH satisfied \textbf{Condition 1}. We can estimate the total DVH with respect to rectum, demonstrated similarly to the average of DVH variations. However, estimation with respect to the bladder deviated from the average because a larger variation of the whole range of the dose-volume affected the estimation. However, when comparing Figs. 4c–d, an obvious inaccuracy was observed for the maximum dose. This inaccuracy occurred because the probability near the boundary was increased slightly by numerical error. Considering detailed time-steps, a higher $a$-value can increase the accuracy of boundary region.
Figure 7 depicts various conditions of $a$, $\Delta T$, and $\mathrm{k}$ conditions. Actually,  $\Delta T \geq 0.1$ can satisfy \textbf{Condition 1}. However, the minimum and maximum dose regions provided obvious error. In contrast, detailed  $\Delta T \leq 0.1$ was able to give accuracy for the minimum and maximum dose regions. The $\Delta T = 0.1$ condition was a balanced solution; it also provided many robust points for evaluation despite its increased  $\Delta T$ resolution.

\begin{figure*}

\includegraphics[width=1.80\columnwidth]{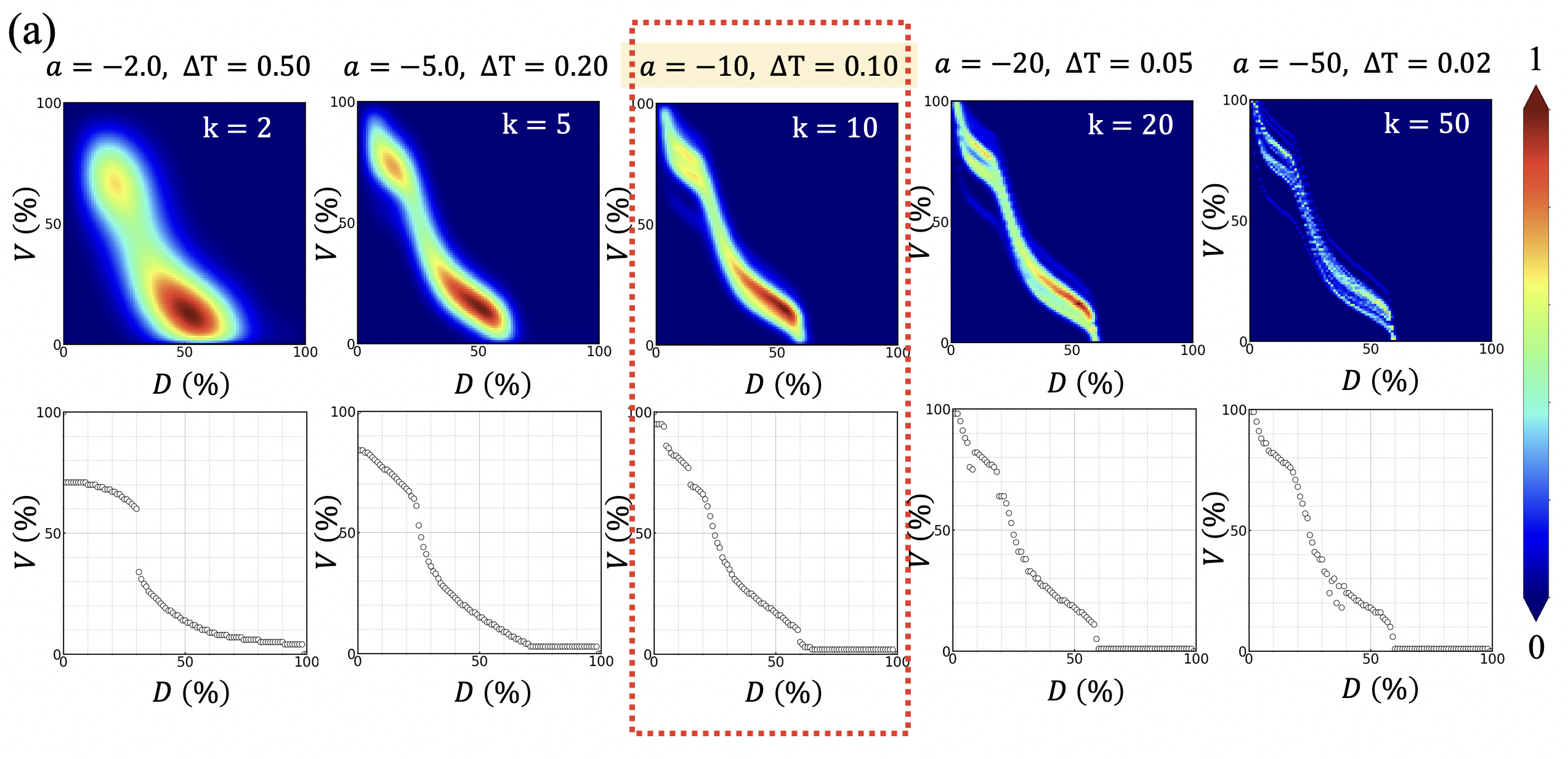}
\includegraphics[width=1.80\columnwidth]{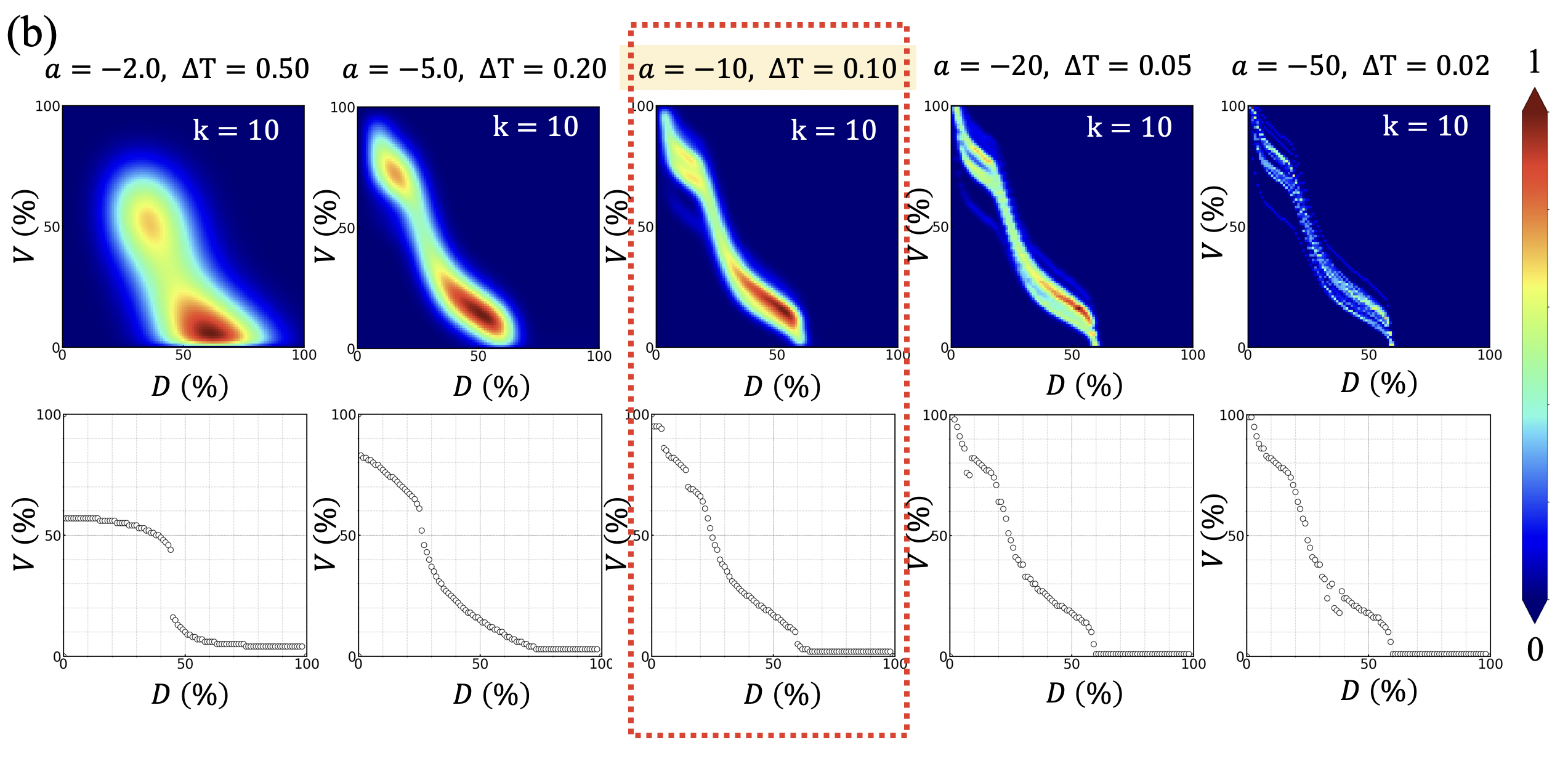}
\caption{\label{fig:wide} 
Comparisons for total DVHs with respect to the  $a$-value and its related parameters of (a) various$ \mathrm{k} \Delta T = 1.0$  and (b) $ \Delta T $ at $ \mathrm{k} = 10 $. We set  $a=-10$ and $\Delta T= 0.1$ as a reasonable standard for this study.}

\end{figure*}

\section{\label{sec:level1} Discussion \protect}

In radiotherapy, an invisible high-energy X-ray or particle is irradiated to the tumor target and the relevant lesions of a patient. To achieve the first planning condition accurately, supporting methods or devices are used such as fixtures for patient posture and image-guided localization. The treatment machinery and the localizing system must have high quality to achieve high-accuracy irradiation by the experts, medical physicists \cite{ref12, ref43, ref44}. However, some uncontrollable factors can raise uncertainty, such as patient involuntary movements, shrunken target volume, changes in organ shape, and shifted organ positions. All can occur in a high dose gradient. Therefore, the same dose to the same volume can entail completely different situations. Moreover, trivial errors of the treatment machinery occur randomly \cite{ref16, ref17, ref18, ref20, ref45, ref46, ref47, ref48}. Therefore, the actual treatment states are never identical. Henriquez et al. \cite{ref26} applied confidence intervals to a DVH, $\alpha$-DVH, and dose expected volume histogram, which enables the estimation of the dose–volume on the true DVH with an appropriate confidence interval. However, such broad-based evaluation ignores convergence of the DVH. Deformable image registration is a groundbreaking technology for estimating the dose distribution in the deformed shape of an organ using the idea of diffeomorphism, which allows the dose distribution to be warped to the initial state of the patient. However, complex processing and algorithms to obtain deformed vector and dose distribution can strongly affect accuracy \cite{ref49, ref50}. Moreover, the evaluation for summation is limited to the initial shape of the organ. Therefore, the true total dose–volume assessment has remained an important unsolved problem.

This study produced one rigorous solution using the concept of optimal transport in TDV space, providing novelty by the superimposition of different states. In this study, it is noteworthy that the bladder received a similarly high dose and the rectum received a greater high dose than the plan. Application of a high dose to the normal organ is clinically important. Our method revealed that the total estimation deviated from the plan and the average dose–volume effect, which would engender the possibility of injury because of high–dose effects. Although dose $D$ and volume $V$ in the TDV space were not simple indices, which superimposed a variable clinical situation of organ dose–volume, we can consider the initial dose–volume relation because of optimal transport from the plan. This fact was verified statistically by Wahl et al. \cite{ref51}, who demonstrated the validity of analytical probabilistic approach in a closed form to combine dose uncertainty with DVH uncertainty. A perturbation of the realistic dose distribution corresponds directly to the DVH uncertainty. Regarding the general interest, Eq. (15) gives a feasible state solution to the many-body problem expressed in probabilistic terms. Eq. (15) can be expanded to more general n–dimensions, which might be useful for finding the optimal solution in a complex combination of stochastic events given certain potential for the initial condition.

In the TDV space, \textbf{Condition 2} appears to be a strong law for probabilistic transportation. More generally, we can apply more random time-dependent metric to the TDV space, but  $\xi$ and $\eta$ are indeterminate, which might be readily apparent because the metric equivalence between $D$ and $V$ is no longer available. Further consideration for the random walk with time-inhomogeneous geodesics was reported by Kuwada \cite{ref52}. Ollivier expanded Ricci-curvature in a discrete space \cite{ref53}, and Lin et al. demonstrated that the lower bound of Ricci curvature is also valid for the Markov chain graph problem \cite{ref54, ref55}. These earlier reports suggest the possibility of certain method that can directly obtain summation of fractionated DVH without \textbf{Condition 2}. However, to solve the problem numerically, we demanded Ricci-flat space considering Euclidean volume/area equivalence in the TDV space.

The limitations of this study are the following. First, the simulation was aimed at the potential distribution for the feasible summation of DVH after transportation in the TDV space. From the optimal transport characteristic (\textbf{Condition 3}), all fractionated DVH and initial planned DVH are connected via the probability elements as coupling set  $\Pi$. Consequently, the transportation of the DVH is viable in the TDV space. The goal of summation of DVH is obtained as the plot line of the highest potential against $D$. To apply it more clinically, the initial distribution of  $\Phi(0,D,V)$  is the key to customization. We set the discretized  $\Phi(0,D,V)$ of treatment and planned DVHs equally. In actuality, the probability elements of treatment DVH are the posterior probability of planned DVH. Therefore, some weight factors between planned and treatment DVHs are required.  

Second, regarding the  $a$-value, this parameter represents an inverse of the diffusion coefficient, which was a hyperparameter in this study. In the fractionated treatment process, organs receive time-varying effects of intra-fractional (during treatment) movements or inter-fractional (between fractions) motions, such as size changes caused by deformation, unexpected intensity modulation attributable to involuntary motions, patient setup uncertainties, and daily treatment machine conditions, causing changes in the radiobiological effects of radiation [43]. These effects are strongly related to the diffusive process. Consequently, the $a$-value should be a function of  $D$ and $V$ rather than a constant,  which is expected to correspond to the patient status and the irradiation delivery process. Therefore, a weighted TDV space for each fractionated treatment should be required. The Bakry–Emery tensor \cite{ref56}, where Ricci curvature with a Hessian is defined for a weighted Riemannian manifold, could provide a space to calculate the TDV transportation with a more appropriate $a$-value as a modified \textbf{Condition 3}, although the condition was beyond the scope of this study. 

Finally, we presume high-fractionated treatment for the feasible total DVH. For this study, we did not consider cases that require a higher dose to a small number of fractions, as with hypo-fractionated stereotactic radiotherapy. Nevertheless, we expect that our concept is still valid because the hypo-fractionated treatment is generally more accurate than high-fractionated treatment with more effort to approximate the patient condition close to the planned condition.

\section{\label{sec:level1} Conclusion \protect}

In radiotherapy, fractionated therapy is common. Clinical estimation is performed using a DVH. However, we typically perform the estimation based on the planned DVH and separately evaluate individual fractionated DVHs. The total DVH considering these fractionated DVHs should be evaluated for the accurate assessment of the treatment effect or side effect, but some difficulty for the superimposition of multiple clinical situations exists. The stochastic consideration can estimate this true total DVH with widely various expectations, but it is extremely difficult to converge to one form. We proposed one solution for the true total DVH using a geometrical approach. In the time–dose–volume space with a time-evolved metric, the DVH potential is transported by the irrotational and incompressible flow. This transportation is optimal transport, thereby allowing consideration of the joint probability set of plan and treatment. The feasible total DVH is obtained as a high potential line against dose. Robust or vulnerable points of the planned DVH can be discussed from geometrical requirements, which are usually underlying the stochastic error.

\section*{Ethical approval}
Experiment protocols for this study were acknowledged by the Institutional Review Board and Independent Ethics Committee of Kansai Medical University (approval number 2022065). All participants gave informed consent by an opt-out approach. Those who rejected were excluded. All methods were conducted in accordance with relevant guidelines and regulations.

\section*{No conflict of interest}
The author has no conflict to disclose.

\section*{Acknowledgments}
This work partly supported by [grant numbers 18K15650, 22H05108, 24K10918] from the Japan Society for the Promotion of Science.

\section*{Appendix}

\appendix
\section{Affine connection coefficient}
Line element $ds$ for TDV space is 
\begin{equation}
	ds^2 = dT^2 + {\xi}^2(T)dD^2 +  {\eta}^2(T)dV^2.
\end{equation}
The Christoffel symbol (Levi–Civita connection coefficient)  $\Gamma^{k}_{ij}$ is represented as
\begin{equation}
	\Gamma^{k}_{ij} = \frac{1}{2} \mathrm{g}^{kl} 
		\left( \partial_i  \mathrm{g}_{lj} + \partial_j  \mathrm{g}_{li} -  \partial_l \mathrm{g}_{ij} \right)
\end{equation}
The geodesics equation is obtained as
\begin{equation}
	\frac{d^2x^k}{ds^2} + \Gamma^{k}_{ij} \frac{dx^i}{ds} \frac{dx^j}{ds} = 0.
\end{equation}
The following Lagrangian $Q$ of
\begin{equation}
	Q = \left(\frac{dT}{ds} \right)^2 + {\xi}^2  \left(\frac{dD}{ds} \right)^2 + {\eta}^2  \left( \frac{dV}{ds} \right)^2
\end{equation}
is rewritten as a Euler–Lagrange equation:
\begin{equation}
	\frac{\partial Q}{\partial x^k} - \frac{d}{ds} \left( \frac{\partial Q}{\partial \left( dx^k/ds\right)} \right) = 0
\end{equation}
Using Eq. (A5) and comparing Eq. (A3)  leads to the connection coefficient.
\begin{equation}
	\begin{split}
		\frac{\partial Q}{\partial T} & - \frac{d}{ds} \left( \frac{\partial Q}{\partial (dT/ds)} \right)  \\
		&= \frac{\partial {\xi}^2}{\partial T} \left( \frac{dD}{ds} \right)^2 +  
			\frac{\partial {\eta}^2}{\partial T} \left( \frac{dV}{ds} \right)^2 - 2\left( \frac{d^2T}{ds^2} \right) \\
		&= 0, \\
	\end{split}
\end{equation}

\begin{equation}
	\begin{split}
		\frac{\partial Q}{\partial D} & - \frac{d}{ds} \left( \frac{\partial Q}{\partial (dD/ds)} \right) \\
		& = \frac{\partial {\xi}^2}{\partial D}  \left( \frac{dD}{ds} \right)^2
		-2\left( 2{\xi} \left( \frac{\partial \xi}{\partial T}  \frac{dT}{ds}\right) \left( \frac{dD}{ds}\right)
		+{\xi}^2 \left( \frac{d^2 D}{ds^2} \right) \right) \\
		& = 0,
	\end{split}
\end{equation}

\begin{equation}
	\begin{split}
		\frac{\partial Q}{\partial V} & - \frac{d}{ds} \left( \frac{\partial Q}{\partial (dV/ds)} \right) \\
		& = \frac{\partial {\eta}^2}{\partial V}  \left( \frac{dV}{ds} \right)^2
		-2\left( 2{\eta} \left( \frac{\partial \eta}{\partial T}  \frac{dT}{ds}\right) \left( \frac{dV}{ds}\right)
		+{\eta}^2 \left( \frac{d^2 V}{ds^2} \right) \right) \\
		& = 0,
	\end{split}
\end{equation}
Therefore, the specific forms of the connection coefficients $\Gamma^{k}_{ij}$ are
\begin{equation}
	\begin{split}
		\Gamma^{T}_{DD} &= -\xi \frac{\partial \xi}{\partial T}, \\
 		\Gamma^{T}_{VV} &= -\eta \frac{\partial \eta}{\partial T}, \\
 		\Gamma^{D}_{TD} &=  \Gamma^{D}_{DT} = \frac{1}{\xi} \frac{\partial \xi}{\partial T}, \\
		\Gamma^{V}_{TV} &=  \Gamma^{V}_{VT} = \frac{1}{\eta}\frac{\partial \eta}{\partial T} , \\
 	\end{split}
\end{equation}

\section{Ricci curvature}
Ricci curvature $R_{ij}$ is
\begin{equation}
	R_{ij} = \partial_{m}  \Gamma^{m}_{ij} - \partial_{j}  \Gamma^{m}_{im} 
				+ \Gamma^{m}_{nm} \Gamma^{n}_{ij} - \Gamma^{m}_{nj} \Gamma^{n}_{im} 
\end{equation}
For each direction, we have,
\begin{equation}
	\begin{split}
	R_{TT} &= \partial_{m}  \Gamma^{m}_{TT} - \partial_{T}  \Gamma^{m}_{Tm} 
				+ \Gamma^{m}_{nm} \Gamma^{n}_{TT} - \Gamma^{m}_{nT} \Gamma^{n}_{Tm} \\
			&= -\partial_{T} \left( \Gamma^{D}_{TD} + \Gamma^{V}_{TV} \right)
				-\left( \Gamma^{D}_{DT}  \Gamma^{D}_{TD} +  \Gamma^{V}_{VT}  \Gamma^{V}_{TV} \right) \\
			&= -\frac{\partial }{\partial T} \left( \frac{\dot{\xi}}{\xi} + \frac{\dot{\eta}}{\eta} \right)
				-\left( \frac{\dot{\xi}}{\xi}\right)^2 -\left( \frac{\dot{\eta}}{\eta}\right)^2,
	\end{split}
\end{equation}
\begin{equation}
	\begin{split}
	R_{DD} &= \partial_{m}  \Gamma^{m}_{DD} - \partial_{D}  \Gamma^{m}_{Dm} 
				+ \Gamma^{m}_{nm} \Gamma^{n}_{DD} - \Gamma^{m}_{nD} \Gamma^{n}_{Dm} \\	
			&=  \partial_{T} \Gamma^{T}_{DD} + \left( \Gamma^{D}_{TD}+\Gamma^{V}_{TV} \right) \Gamma^{T}_{DD} \\
				&\quad - \left( \Gamma^{D}_{TD} \Gamma^{T}_{DD} + \Gamma^{T}_{DD} \Gamma^{D}_{DT} \right) \\
			&= \partial_{T} \Gamma^{T}_{DD} + \left( \Gamma^{V}_{TV} - \Gamma^{D}_{DT} \right) \Gamma^{T}_{DD} \\
			&= \frac{\partial}{\partial T}  \left( -\xi \dot{\xi} \right) 
				+ \left( \frac{\dot{\eta}}{\eta} - \frac{\dot{\xi}}{\xi} \right) \left(-\xi \dot{\xi} \right) \\
			&= -{\xi}^2 \left( \frac{\ddot{\xi}}{\xi} + \frac{\dot{\xi}\dot{\eta}}{\xi \eta} \right),
	\end{split}
\end{equation}
\begin{equation}
	\begin{split}
	R_{VV} &= \partial_{m}  \Gamma^{m}_{VV} - \partial_{V}  \Gamma^{m}_{Vm} 
				+ \Gamma^{m}_{nm} \Gamma^{n}_{VV} - \Gamma^{m}_{nV} \Gamma^{n}_{Vm} \\	
			&=  \partial_{T} \Gamma^{T}_{VV} + \left( \Gamma^{V}_{TV}+\Gamma^{D}_{TD} \right) \Gamma^{T}_{VV} \\
				&\quad - \left( \Gamma^{V}_{TV} \Gamma^{T}_{VV} + \Gamma^{T}_{VV} \Gamma^{V}_{VT} \right) \\
			&= \partial_{T} \Gamma^{T}_{VV} + \left( \Gamma^{D}_{TD} - \Gamma^{V}_{VT} \right) \Gamma^{T}_{VV} \\
			&= \frac{\partial}{\partial T}  \left( -\eta \dot{\eta} \right) 
				+ \left( \frac{\dot{\xi}}{\xi} - \frac{\dot{\eta}}{\eta} \right) \left(-\eta \dot{\eta} \right) \\
			&= -{\eta}^2 \left( \frac{\ddot{\eta}}{\eta} + \frac{\dot{\xi}\dot{\eta}}{\xi \eta} \right).
	\end{split}
\end{equation}

$R_{TD} = R_{DT} = 0$, $R_{TV} = R_{VT} = 0$, and $R_{DV} = R_{VD} = 0$ are in the same manner.

Using $R_{DD} = 0$ and $R_{VV} = 0$ condition with $\xi \neq 0$ and $\eta \neq 0$, the following is obtained.
\begin{equation}
	\frac{\ddot{\xi}}{\xi} = \frac{\ddot{\eta}}{\eta} = -\frac{\dot{\xi}\dot{\eta}}{\xi \eta}.
\end{equation}
This suggests we should solve the following partial differential equation
\begin{equation}
	\frac{\partial^2}{\partial T^2} \left( \xi \eta \right) = 0.
\end{equation}
Here, $a(D,V)$ and $b(D,V)$ are constant coefficients for $T$-direction. Using them in the solution of Eq.(B6), then, the following equation holds.
\begin{equation}
	\xi\eta = a(D,V)T+b(D,V)
\end{equation}
However, the definition of $\xi:=\xi(T)$ and $\eta:= \eta(T)$ requires $a$ and $b$ are constants.

\clearpage

\section*{  }

\bibliography{MS_TDV_Anetai_Kotoku_vtex_arXiv.bib}

\end{document}